\documentstyle[aps,floats,epsf]{revtex}
\begin{document}

\baselineskip .3in
\begin{center}
{\Large {\bf The spin-Peierls instability in spin 1/2 XY chain}} 

{\Large {\bf in the non adiabatic limit}}
\end{center}

\vspace {.1in}

\begin{center}
Shreekantha Sil 
\end{center}

\begin{center}
Institut f$\ddot{u}$r Theoretische Physik, Universit$\ddot{a}$t 
zu K$\ddot{o}$ln,\\  
Z$\ddot{u}$lpicher Str. 77, D--50937 K$\ddot{o}$ln, Germany 
\end{center}

\vspace {.3in}
\begin{center}
{\bf Abstract}
\end{center}

 The spin-Peierls instability in spin 1/2 XY chain 
coupled to dispersionless phonons of frequency $\omega$ has been studied in the 
nonadiabatic limit. We have chosen the 
Lang-Firsov variational wave function for the phonon subsystem to obtain an 
effective spin Hamiltonian. The effective spin Hamiltonian is then solved 
in the framework of mean-field approximation. We observed a dimerized phase 
when g is less than a critical value and an anti-ferromagnetic phase when 
it is greater than a critical value . The variation of lattice 
distortion, dimerized 
order parameter and energy gap with spin phonon coupling parameter 
has also been investigated here.  

\vspace{.2in}

\noindent {\bf 1.  Introduction}

The discovery of the spin Peierls transition in $CuGeO_3$ \cite{hase} 
has sparked an 
intense effort to study the properties of this quasi one-dimensional 
magneto-elastic system where the coupling of the magnetic to the 
lattice degrees of freedom 
leads to a phase transition into a dimerized phase. This magneto elastic 
transition is due to the competition between the gain in magnetic energy due 
to dimerization and the loss in the elastic energy of the lattice distortion. 
Recently quite a large number of experimental \cite{palm,lus,ren,buch} 
and theoretical \cite{cross,fuji,nak,got,cas,rie,bou1,bou2} works have been 
performed to investigate the various aspects of the spin Peierls systems. 
Except a few \cite{car,uh} most of the theoretical investigations rely on the 
adiabatic treatment of the 
phonons. In adiabatic approaches one assumes that the phonons responsible 
for the distortion have low energy with respect to the characteristic energies 
for the spin  systems (e.g. the gap=$\Delta$). 
 The experimental evidence for 
 $CuGeO_3$  indicates that the application of the adiabatic approximation 
to these system is not sufficient. Regnault et al. \cite{reg} 
investigated the spin 
dynamics of spin Peierls system $CuGeO_3$ by inelastic neutron scattering. 
Their result confirmed the existence of a gap ($\Delta$) in the magnetic 
excitations at  $\Delta = 2meV = 23 K$ and the Heisenberg exchange parameter 
was considered to be $J_1 = 10.6meV = 115 K $. Braden et al. \cite{bar} 
found by symmetry that four optical 
phonons are possible candidates for the spin Peierls distortion in $CuGeO_3$. 
Among these four modes two, one of energy $330^oK$ and other of energy 
$150^oK$ are experimentally found to be the suitable candidates for the 
spin Peierls distortion. In both cases we find that the phonon frequency 
is larger than $J$ as well as $\Delta$. 
When $\omega > \Delta$ the spin phonon interaction is unretarded and 
 non-adiabatic 
effects or  quantum lattice fluctuations become important. Fradkin 
and Hirsch \cite{frad} considered the Su-Schrieffer-Heeger model \cite{su} 
of electron-phonon 
interaction for spin-less fermions and spin $\frac{1}{2}$ electrons in one 
dimension 
to investigate the stability of the Peierls-dimerized ground state against 
 quantum fluctuations. In this work they have shown by renormalization group
 arguments and by quantum Monte Carlo simulation that for spin-less fermions   
quantum lattice fluctuations destroy the long-range 
dimerization order when the fermion-phonon coupling constant is small and 
predicted a transition from an undimerized ground state to a dimerized phase 
when the 
fermion-phonon interaction is larger than a critical value. Campbell and 
Bishop \cite{cam} independently confirmed the findings of 
Fradkin and Hirsch. Recently 
Caron and Moukouri \cite{car} studied the XY spin chain coupled to 
dispersionless phonons by  the density matrix renormalisation group 
(DMRG) method and showed that  
quantum fluctuations reduce the spin Peierls gap and even destroy 
the dimerization when the phonon frequency gets appreciably larger 
than the gap.    

The coupled spin-phonon system for all values of the coupling 
parameter is a very difficult problem. However, one may gain considerable 
insight into the stability problem by simply considering  limiting 
situations. By analyzing the stability of these limits a qualitative picture 
of possible phases (or the ground state ) will emerge.  
 In the present work we will 
investigate the effect of spin-phonon interaction in the non-adiabatic limit 
i.e. when phonons are certainly fast. In this case we will treat the phonons 
as fast variables and derive an effective interacting fermion model. 

      We propose to study an XY spin chain whose magnetic interaction 
depends on the bond length. The reason for this study is two fold 
i) The undeformed spin chain Hamiltonian can be solved exactly. 

ii) The model contains the essential elements for the spin-Peierls 
transition, that is coupling to intermolecular motion. 
   In the future it will be of interest to study the Heisenberg chain with 
next-nearest neighbor frustration term to make the model more realistic to 
the inorganic spin-Peierls systems. 

   The paper is organized as follows. In section 2 the variational ground 
state energy of the XY model in presence of spin-phonon interaction is 
determined 
using suitable phonon states and mean field approximation taking into account 
the dimerization as well as anti-ferromagnetic ordering. The results of 
numerical solution and its implication are discussed in section 3.

\vspace{.3in}
   
{\bf 2.  Formulation}

We start with the XY model in presence of spin phonon interaction on a 
linear chain:

\begin{equation}
H = \sum_l [(J + g (b_l^ \dagger + b_l - b_{l+1}^ \dagger -b_{l+1})]
(S_l^X S_{l+1}^X
+ S_l ^ Y S_{l+1} ^ Y) + \omega  \sum_l b_l ^ \dagger b_l
\end{equation}
where $l$ denotes the site index of the $N$ site linear chain, 
$S_l^X$ and $S_l^Y$ are components of
the local 
X-Y spin of value $\frac{1}{2}$, $b_l (b_l^\dagger)$ is the annihilation 
(creation)
operator for a vibration of molecule at site $l$ and $J$ is the magnetic 
exchange interaction between the nearest neighbor spins. Here $\omega$ accounts for the dispersionless vibrational spectra for molecular motion along the 
chain direction and g is the spin-phonon interaction \cite{bray}. 

 We transform the spin operator to a spinless pseudo fermion 
representation using the Jordan Wigner transformation \cite{jor} 
\begin{eqnarray}
S_l^X+iS_l^Y &=& S_l^+ = exp( -i \pi \sum_j ^{l-1} d_j^\dagger d_j) 
d_l^\dagger , \\
S_l^X-i S_l^Y &=& S_l^- = exp(i \pi \sum_j^{l-1} d_j^\dagger d_j) d_l , \\
S_l^z &=& \frac{1}{2} + d_l^\dagger d_l 
\end{eqnarray}
to make use of the growing understanding of one dimensional Fermi system.

After the Jordan Wigner transformation the Hamiltonian (1) can be written 
in terms of fermion operators $d_l^\dagger$ and $d_l$ as 
\begin{eqnarray}
H= \frac{1}{2} J \sum_l P_l +\frac{1}{ 
2} g \sum_l (b_l^\dagger + b_l)(P_l - P_{l-1}) + \omega \sum_l b_l^\dagger b_l
\end{eqnarray}
where 
\begin{equation}
P_l = d_l^\dagger d_{l+1} + d_{l+1}^\dagger d_l .
\end{equation}

In the adiabatic approximation the spin-phonon interaction deforms the lattice 
to undergo the Peierls instability. To take into account the lattice 
distortion due to spin phonon coupling in our case we choose a variational 
wave function   
$|\psi \rangle _ {ph} = U |0 \rangle $ with

\begin{eqnarray}
U= exp(\frac{\lambda}{2 \omega}\sum_l (b_l^\dagger - b_l)(P_l - P_{l-1}))
\end{eqnarray}
for the phonon subsystem, where $|0 \rangle$ is the zero phonon state  
and U describes a  
modified Lang-Firsov transformation \cite{lang,das}. In this formalism the 
effective 
fermion Hamiltonian is written as 
\begin{equation}
H_{eff} = \langle 0| H_T |0 \rangle ,
\end{equation}
with 
\begin{eqnarray}
 H_T &=&  U^{-1} H U \nonumber \\
     &=& \frac{J}{2} \sum_l U^{-1} P_l U
+ \frac{g-\lambda}{2} \sum_l (b_l^\dagger + b_l)(P_l - P_{l-1})  \nonumber \\
&-& 4 g'^2 \sum_l n_l + 4 g'^2 \sum_l n_l n_{l+1} + g'^2 \sum (1- 2 n_l)
(d_{l-1}^\dagger d_{l+1} + d_{l+1}^\dagger d_{l-1})  \nonumber  \\
&+& \omega \sum_l b_l^\dagger b_l 
\end{eqnarray}
where,
\begin{eqnarray}
n_l &=& d_l^\dagger d_l ,\\
g'^2 &=& (\frac{g \lambda}{2 \omega} - \frac{\lambda^2}{4 \omega}) .
\end{eqnarray}
In the above equation $b_l$ and $b_l^\dagger$ is the creation and annihilation 
operator for the phonon system vibrating about the displaced equilibrium 
position $ \frac{\lambda}{\omega}(P_l - P_{l-1})$ of the lattice. Clearly, 
$\lambda$  is proportional to a lattice displacement created by the  
spin-phonon 
interaction which has to be determined variationally. When $\lambda = g$ the 
transformation is exactly the Lang-Firsov \cite{lang} transformation where 
the fermion-phonon term is diagonalised exactly and the fermion 
hopping term is 
renormalized by dressed phonons. To obtain an effective fermionic Hamiltonian 
we take the average over the zero phonon state of the transformed phonon 
subsystem and neglect the terms of the order of $\frac{\lambda^4}{\omega^4}$  
and higher. In this approximation the effective Hamiltonian is 
\begin{eqnarray}
 H_{eff}(\lambda) &=& \frac{J}{2}(1- \frac{3 \lambda^2}{4 \omega^2}) \sum_l 
(d_l^\dagger d_{l+1} + d_{l+1}^\dagger d_l) + \frac{J \lambda ^2}{8 \omega ^2}
 \sum_l (d_l ^ \dagger d_{l+3} + d_{l+3}^\dagger d_l)  \nonumber \\
&-& 4 g'^2 \sum_l n_l + 4 g'^2 \sum_l n_l n_{l+1} + g'^2 \sum (1- 2 n_l)
(d_{l-1}^\dagger d_{l+1} + d_{l+1}^\dagger d_{l-1}) \nonumber \\
&+&  O(\lambda^4) .
\end{eqnarray}

Now, the above effective hamiltonian (12) is complicated enough to solve 
it exactly. Therefore one has to look for the approximate methods.  
 We calculate the ground state energy $E_{eff}(\lambda)$ of the effective 
Hamiltonian (13) in the framework of mean field theory. We assume solutions 
which break the symmetry between even and odd sites with respect to the 
number of fermions on the site and to the corresponding hopping probability. 
Both 
of these kinds of oder open a gap at the Fermi momentum  at half filling (ie. 
when the total spin $M_z$ is zero) and lower the ground state energy.  
We will consider four variational parameters such as $n_e$ , $n_o$, $h_e$ 
and $h_o$ (
where, e implies even sites and o implies odd sites). All these variables are 
not independent because they are subject to the fermion number conservation 
constraint 
\begin{equation} 
n=  \frac{< d_{2l}^ \dagger d_{2l} + d_{2l+1}^ \dagger d_{2l+1}>}{2} = 
\frac{(n_e + n_o)}{2}
\end{equation}
For half filling ($M_z=0$) $n= \frac{1}{2}$. The three remaining variational
 parameters are then the 
anti-ferromagnetic order parameter 
\begin{equation}
m = \frac{< d_{2l}^ \dagger d_{2l} - d_{2l+1}^ \dagger d_{2l+1}>}{2} = 
\frac{(n_e - n_o)}{2},
\end{equation}
the dimerized order parameter 
\begin{equation}
\gamma =  \frac{< d_{2l}^ \dagger d_{2l+1} - d_{2l+1}^ \dagger d_{2l+2}>}
{2} = \frac{(h_e - h_o)}{2}
\end{equation} 
and average hopping probability
\begin{equation}
h =   \frac{< d_{2l}^ \dagger d_{2l+1} + d_{2l+1}^ \dagger d_{2l+2}>}
{2} = \frac{(h_e + h_o)}{2}
\end{equation} 
 Here $< -- >$ implies the expectation value over the ground state. 
Within the limitation of Hartree-Fock 
approximation our effective Hamiltonian can be written as
\begin{eqnarray}
 H_{eff} &=& \frac{J}{2}(1- \frac{3 \lambda^2}{4 \omega^2} 
) \sum_l 
(d_l^\dagger d_{l+1} + d_{l+1}^\dagger d_l) + \frac{J \lambda ^2}{8 \omega ^2}
 \sum_l (d_l ^ \dagger d_{l+3} + d_{l+3}^\dagger d_l)  \nonumber \\
&+&  8 g'^2 \gamma \sum_l (-1)^l (d_l^ \dagger d_{l+1} + d_{l+1}^ \dagger d_l)
 + 8 g'^2 \sum (\frac{1}{2} - (-1)^l m) n_l) \nonumber \\
& - & 4 g'^2 (\frac{1}{4} - m^2 - 2 \gamma^2) N - 2 g'^2 N
+  O(\frac{\lambda^4}{\omega ^4})
\end{eqnarray}
 To diagonalise the Hamiltonian (18) we transform the operators from 
coordinate space to momentum space 
\begin{eqnarray} 
c_j^\dagger & = & \frac{1}{\sqrt{N}} \sum_k c_k^ \dagger e^{i k j} \\
c_j & = & \frac{1}{\sqrt{N}} \sum_k c_k e^{-i k j}
\end{eqnarray}
Due  to the reduced symmetry each $k$ state  
is coupled to the state $k + \pi$. So it is convenient to write the 
Hamiltonian in the reduced zone $ - \frac{\pi}{2} $ to $ \frac{\pi}{2} $ and 
label the states by  band indices l and u. In this representation the 
Hamiltonian is a two band Hamiltonian where the band $l$ and $u$ are 
coupled to each $k$.   
\begin{eqnarray}
H_{eff} &=& \sum_k \alpha_k^l d_k^{l \dagger} d_k^l + \sum_k \alpha_k ^u 
d_k^{u \dagger} d_k^u + \sum_k \beta_k d_k^ {l \dagger} d_k^u + \sum_k
\beta_k^* d_k ^{u \dagger} d_k^l \nonumber  \\
& - & 4 g'^2 (\frac{1}{4} - m^2 - 2 \gamma^2) N  
+ O(\frac{\lambda^4}{\omega ^4})
\end{eqnarray}
 where,
\begin{eqnarray}
\alpha_k ^l &= & J(1 - \frac{3 \lambda ^2}{4 \omega^2}) cos(k) + 
\frac{2J \lambda^2} {8 \omega^2} cos(3k) \\
\alpha_k ^u &= & - J(1 - \frac{3 \lambda ^2}{4 \omega^2}) cos(k) - 
\frac{2J \lambda^2} {8 \omega^2} cos(3k) \\
\beta_k &=& - 8 g'^2 (m - 2i \gamma sin(k)) 
\end{eqnarray}
 
We diagonalise this Hamiltonian by a Bogoliubov Valatin transformation and 
obtain   
\begin{eqnarray}
\frac{H_{eff}}{\omega} & = & \sum_k E_k^l a_k^ \dagger a_k + \sum_k E_k^u b_k^ \dagger b_k 
\nonumber \\
& - & 4 g'^2 (\frac{1}{4} - m^2 - 2 \gamma^2) N 
\end{eqnarray}
with upper and lower band  
\begin{equation}
E_k ^{u/l} = \pm \sqrt{ \frac{\alpha_k ^{l 2}}{\omega^2} 
+ 64 \frac{g'^4 m^2}{\omega^2}
 + \frac{256 g'^4 \gamma^2 sin(k)^2}{\omega^2}}
\end{equation}

From the equation (25) it is clear that the energy spectrum has been split 
into two separate bands ( for non zero $m$ or $\gamma$) characterized 
by the Bogoliubov transformed creation 
(annihilation) operators $a_k^ \dagger (a_k)$ and $ b_k^ \dagger (b_k)$. 
For half-filling ( $M_z = 0$) the lower band is completely filled in 
 the ground state . 
 We will take the expectation value of the equations (14) (15) over the 
ground state to obtain  a set of self-consistent equations of       
\begin{equation}
m= 8 \frac{g'^2 m}{\omega N} \sum_{k = - \pi /2} ^ {k= \pi/2}   
\frac{1}{|E_k^l|}
\end{equation}
and 
\begin{equation}
\gamma = 16 \frac{g'^2 \gamma}{\omega N} \sum_{k = - \pi /2} ^ {k= \pi/2}   
\frac{sin(k)^2}{|E_k^l|}
\end{equation}
If we retain terms up to the order of $\frac{g^2}{\omega^2}$ the 
integrals involving the 
equations (26) and (27) can be expressed in terms of elliptic functions 
of the first kind $( K(\nu))$ and the second kind $( E(\nu))$ to give 
\begin{equation}
m  = \frac{8 g'^2 \sqrt{\nu}~ m }{\pi J (1- \frac{3 \lambda^2}{2 \omega^2})} 
K(\nu)   
\end{equation}
with $ \nu = \frac{1}{1 + \frac{64 g'^4 m^2}{J^2(1-3 \lambda^2/(2 \omega^2))}}$
 and
\begin{equation}
\gamma  = -\frac{32 g'^2  \gamma }{\pi J (1- \frac{3 \lambda^2}
{2 \omega^2})} \frac{d}{d \nu'} E(\nu')   
\end{equation}
and $ \nu' = 1 - \frac{256 g'^4 \gamma^2}{J^2(1-3 \lambda^2/(2 \omega^2))}$. 
 For small $m$ or $\gamma$ the equations provide asymptotic expressions for 
1) the antiferromagnetic and 2) the dimerized phase. 

1) The antiferromagnetic phase ($\gamma =0 ; m \neq 0$)
   
\begin{eqnarray}
\gamma &=& 0  ; \nonumber \\
m &=&  p_j  exp(-  \pi p_j (1 - \frac{1.38629}{p_j}) 
\end{eqnarray}

2) The dimerized phase ($m=0 ; \gamma \neq 0$)

\begin{eqnarray}
m &=& 0 ;  \nonumber \\
\gamma &=&  \frac{pj}{2}  exp(- \frac{ \pi pj}{2} (1 - \frac{0.77308}{pj}) 
\end{eqnarray}
where $p_j = J \frac{1 - 3 \lambda^2/(2 \omega^2)}{8 g'^2} $

When $m$ or $\gamma$ is large we solve the equation (26) or (27) numerically 
to get the anti-ferromagnetic or dimerized order parameter. 

 Finally we minimize the ground state energy ($E_G(\lambda)$) of the 
Hamiltonian $H_{eff}$ with respect to $\lambda$ to get $\lambda$ and 
the ground state energy of the system.   

\vspace{.3in}

{\bf 3.  Results and discussion}

We have computed the ground state energy, the lattice distortion and 
the  energy gap for the XY system in presence of spin-phonon 
interaction when phonon frequency is certainly fast. The results of 
our calculation shows some distinguished features in the phase diagram  
which is absent when phonons are treated adiabatically. In the adiabatic 
approach the spin-phonon coupling to the XY spin system always gives rise 
to dimerized phase where as our nonadiabatic approach predicts spin-liquid,
dimerized and antiferromagnetic phase depending on the values of
$\frac{g}{\omega}$ and $\frac{J}{\omega}$ ratio.     
 Here we present the result for two 
values for exchange 
interaction $\frac{J}{\omega}$ = 0.2 and 0.4. In our calculation we have 
neglected the terms of the order $\frac{\lambda^4}{\omega^4}$ or higher. So 
we do not extend our calculation for large spin-phonon coupling 
$\frac{g}{\omega}$ and 
 confine ourselves to $\frac{g}{\omega} \le 0.5$ . In figure 1. we have 
shown the variation of lowest energy in the anti-ferromagnetic as well as 
dimerized phase to determine the ground state of the system. For $\frac{J}
{\omega} = 0.2$ we find that the dimerized phase represents the ground state 
when $\frac{g}{\omega} <.46$ and anti-ferromagnetic phase becomes the ground 
state when $\frac{g}{\omega} > 0.46$. The critical value for dimerized to 
anti-ferromagnetic phase transition increases with the increase of $\frac{J}{
\omega}$. For J=0.4 the above mentioned critical value of $\frac{g}{\omega}$ 
is higher than $0.5$ and it is not shown in our plot because 
we have confined our investigations within $\frac{g}{\omega}=0.5$.

In figure 2. we show the 
variation of dimerized order parameter $\gamma$ with $\frac{g}{\omega}$ for 
$\frac{J}{\omega} = 0.2$ and $0.4$. The figure shows that 
when $\frac{g}{\omega}$
is less than 0.1 the dimerized order parameter is almost zero. From the 
equation (31) we obtain $\gamma  \sim  10^{-20}$ at $\frac{g}{\omega} =0.1$. In the 
mean-field calculation we neglect quantum fluctuations and always find a 
nonzero value of the order parameter. However, the quantum fluctuation due to 
spin excitation would have drastic repercussions on the value of 
$\gamma$ and can destroy 
this small value of $\gamma$ to give a spin liquid phase. So one may expect a 
critical spin-phonon coupling ($\frac{g_c}{\omega}$) for the on set of 
dimerization. Caron and Moukouri \cite{car} have shown 
this existence of critical 
spin - phonon coupling ($\frac{g_c}{\omega}$) through the density matrix 
renormalization group calculation where quantum fluctuation are taken 
into account rigorously. It is also 
evident from our figure 2. that the dimerization order parameter decreases 
with the increase of $\frac{J}{\omega}$ which means that we expect a larger 
$\frac{g_c}{\omega}$ for $\frac{J}{\omega} = 0.4$.            

Due to Peierls instability we have two types of bond length one is greater 
than the unperturbed lattice constant and other is less than the unperturbed 
lattice constant. We represent this lattice deformation by $
<b_l ^ \dagger + b_l -b_{l+1}^\dagger -b_{l+1}> = (-1)^l \delta$. In 
figure 3. we plot the variation of $\delta = 2 \lambda \gamma$  with respect 
to $\frac{g}{\omega}$ for $\frac{J}{\omega} = 0.2 $ and $0.4$ in the dimerized 
phase. It is seen from the figure that for small $\frac{g}{\omega}$ ratio 
the lattice distortion is almost zero and it increases with the increase of
$\frac{g}{\omega}$ or with the decrease of $\frac{J}{\omega}$.   

 We have also investigated the excitation energy gap ($\Delta$) which is 
obtained by calculating  the lowest 
 excited state energy  (ie. lowest energy of the upper band ) 
. In the figure 4. we plot the variation 
of $\Delta$ with respect to $\frac{g}{\omega}$ for 
$\frac{J}{\omega} = 0.2$ and $0.4$. Like the dimerization order 
parameter, $\Delta$ 
is also very small for small spin-phonon coupling. The result also points to 
the possibility of gapless spectra for $\frac{g}{\omega} < \frac{g_c}{\omega}$
and gapfull spectra for $\frac{g}{\omega} > \frac{g_c}{\omega}$. 
 
In figure 5. we show the variation of the critical value $\frac{g'_c}{\omega}$
for dimerized to antiferromagnetic phase transition with respect to $\frac{J}{
\omega}$. We observe a decrease of $\frac{g'_c}{\omega}$ with the decrease 
of $\frac{J}{\omega}$ and in the limit $\frac{J}{\omega} \rightarrow 0$ 
the system 
shows an antiferromagnetic order for any finite spin-phonon coupling. 
This feature 
is significantly different if the spin-phonon coupling 
is treated adiabatically where the ground state of XY model 
with spin-lattice interaction always represents dimerized phase. Our results 
signifies that the quantum correction may play a very important role in 
determining the phases of the ground state of the  spin-Peierls systems.   
  
In summary, we have developed a nonadiabatic approach for the interacting 
spin-phonon problem. We have chosen the Lang-Firsov variational wave 
function to integrate out the phonon degrees of freedom and obtained 
an effective spinless fermionic hamiltonian which is solved in the frame 
work of mean field approximation to calculate the ground state energy,
the minimum excitation energy gap and the lattice distortion developed in 
the ground state. Our investigation indicates two types phase transition 
one is from spin liquid phase to dimerized phase and another is from 
dimerized phase to antiferromagnetic phase as we vary the spin-phonon coupling 
from a very low value. However, these phase transitions are absent if 
we neglect the quantum lattice fluctuations and treat the problem in the 
adiabatic approximation. So it is evident that the behavior of spin-Peierls 
system is significantly modified by quantum lattice fluctuations and an 
intense investigation is required to explore the effect of nonadiabatic 
corrections to the spin-Peierls systems. 
For a realistic calculation of inorganic spin-Peierls 
system like $CuGeO_3$  one has to study the Heisenberg spin chain 
with next nearest neighbor frustration . 

\vspace{.3in}

{\bf Acknowledgments}

    The author acknowledges the helpful discussions with 
E. M$\ddot{u}$ller-Hartmann, G. Bouzerar and 
 G. Uhrig.  Research was performed within the program of the 
Sonderforschungsbereich 341 supported by the Deutsche Forschungsgemeinschaft 
(DFG).

\newpage

{\bf Figure captions}

{\bf Figure 1.} Variation of the minimum energy (in the unit of the energy 
of the phonon) with respect to the spin-phonon 
coupling ($\frac{g}{\omega}$) in the antiferromagnetic 
phase for J=0.2 (curve $a$) and J= 0.4 (curve $c$) and in the 
dimerized phase for J=0.2 (curve $b$) and J=0.4 (curve $d$). 

{\bf Figure 2.} Variation of the dimerized order parameter $\gamma$ with 
respect to $\frac{g}{\omega}$ for $J= 0.2$ (curve $a$) and $J=0.4$ (curve $b$) 
 in the dimerized phase. 

{\bf Figure 3.} Variation of the lattice distortion $\delta$ with respect to 
$\frac{g}{\omega}$ in the dimerized phase for $J=0.2$ (curve $a$) and 
$J=0.4$ (curve $b$). 

{\bf Figure 4.} The  excitation energy gap $\Delta$ as a function 
of $\frac{g}{\omega}$ 
in the dimerized phase for $J=0.2$ (curve $a$) and $J=0.4$ (curve $b$).  

{\bf Figure 5.} Variation of the critical value $\frac{g'_c}{\omega}$
for dimerized to antiferromagnetic phase transition with respect to 
$\frac{J}{ \omega}$.

\begin{figure}
\centerline{
\epsfxsize=18.0cm
\epsffile{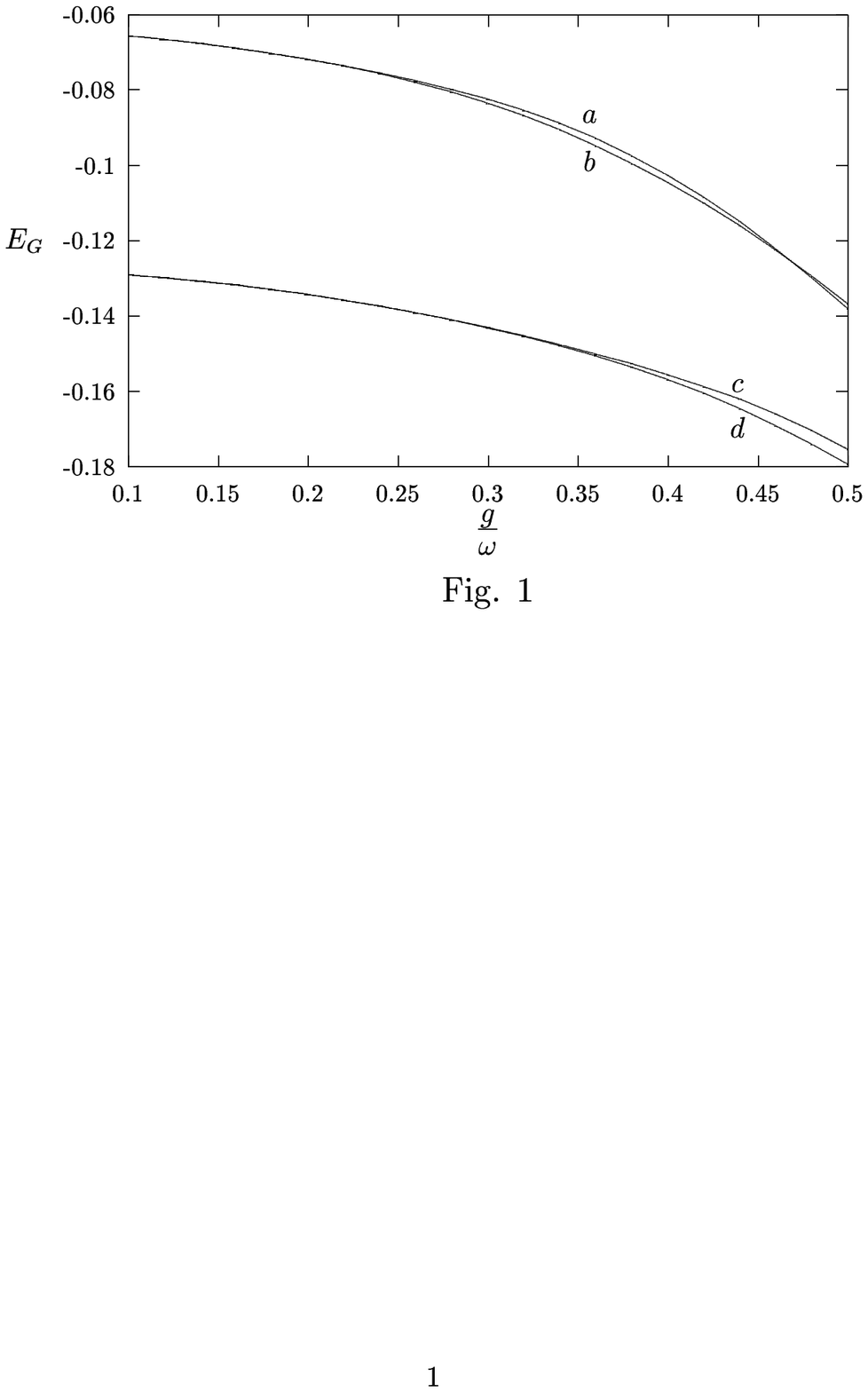}}
%\label{fig1}
\end{figure} 

\begin{figure}
\centerline{
\epsfxsize=18.0cm
\epsffile{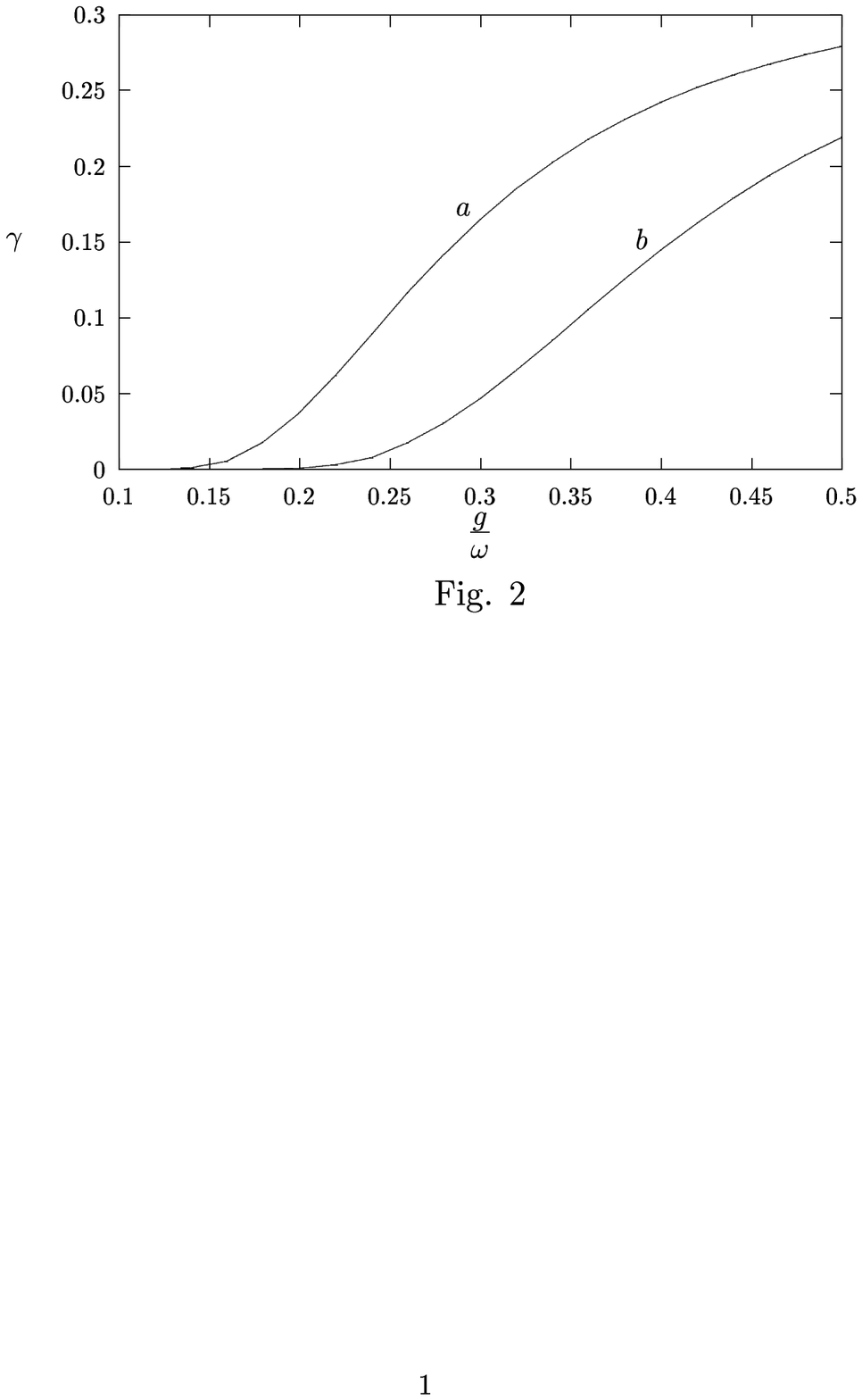}}
%\label{fig2}
\end{figure}

\begin{figure}
\epsfxsize=18.0cm
\epsffile{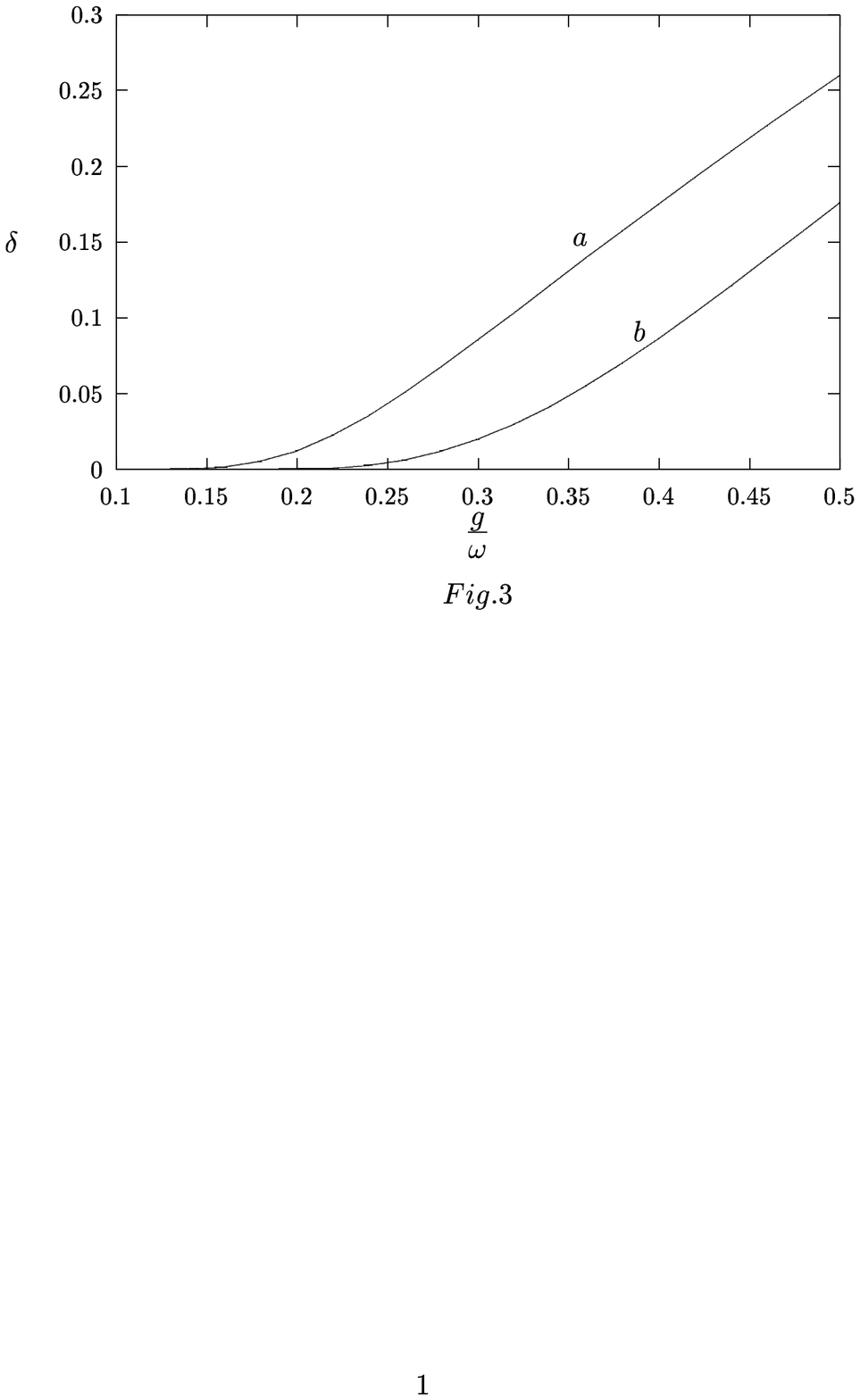}
%\label{fig3.ps}
\end{figure}

\begin{figure}
\epsfxsize=18.0cm
\epsffile{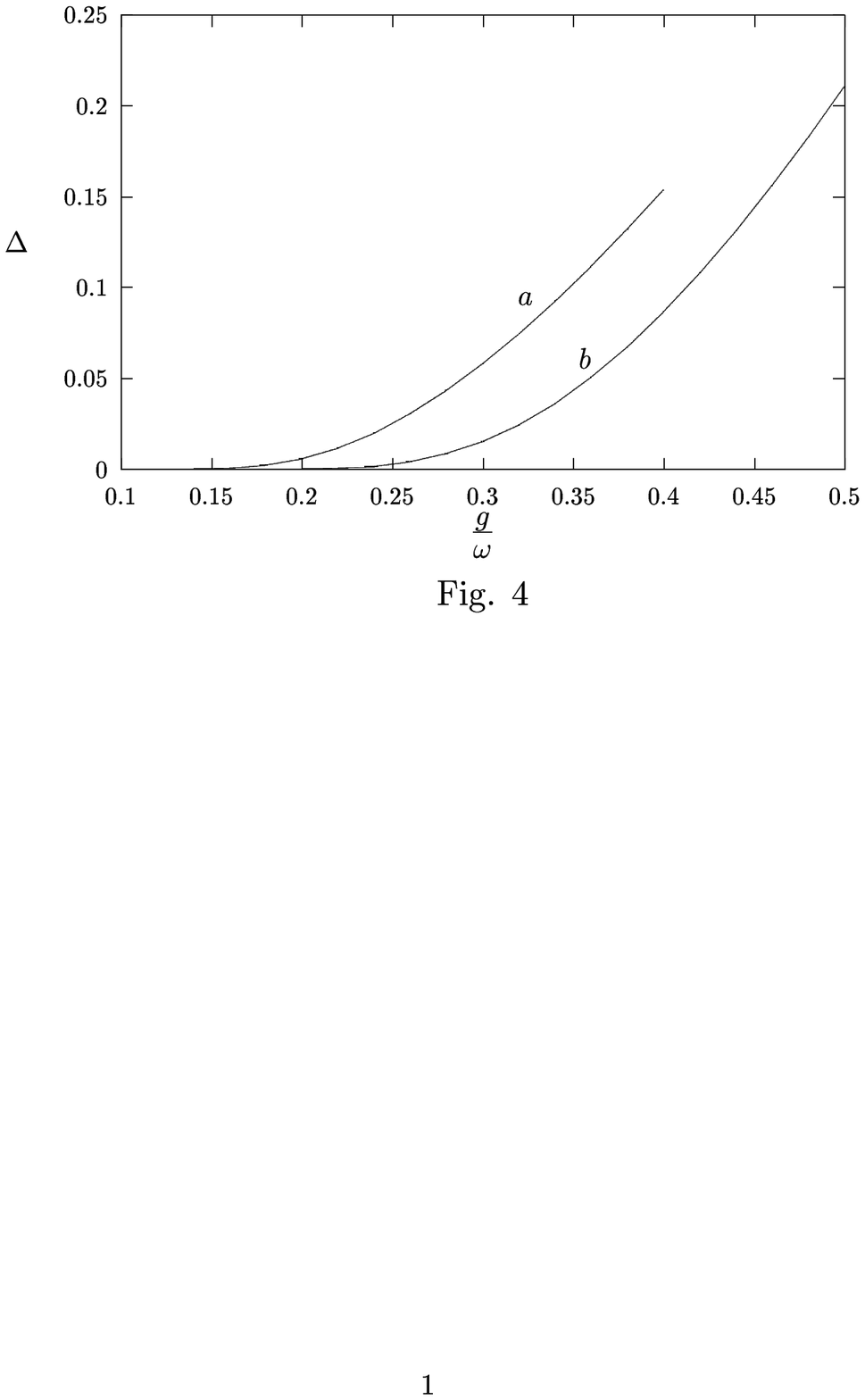}
%\label{fig4}
\end{figure} 

\begin{figure}
\epsfxsize=18.0cm
\epsffile{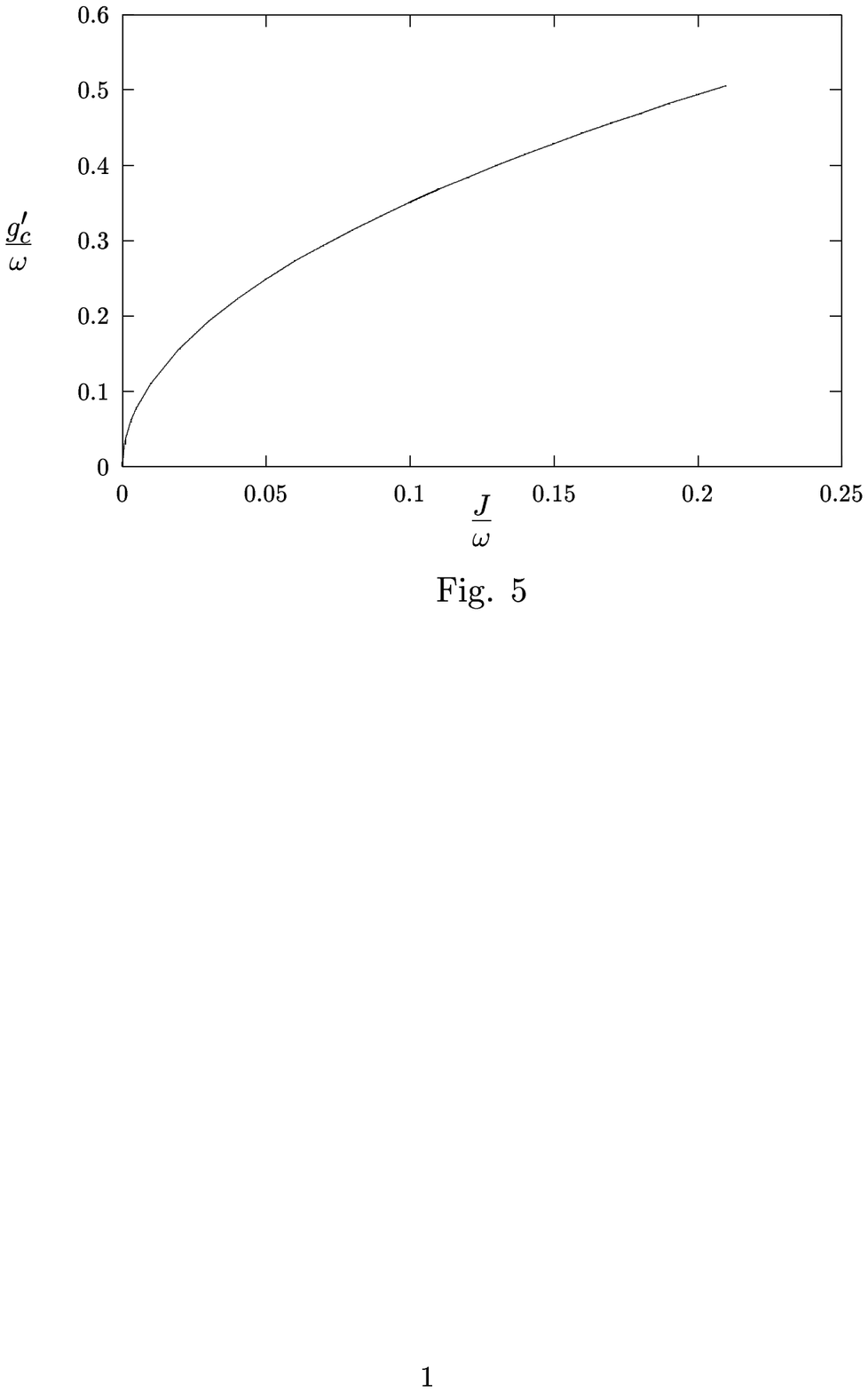}
%\label{fig5}
\end{figure} 


\vspace{.3in}

\begin{references}

\bibitem{hase} M. Hase, I. Terasaki and K. Uchinokura, Phys. Rev. Lett. 
{\bf 70}, 3651 (1993)

\bibitem{palm} W. Palme et al. J. Appl. Phys. {\bf 79}, 5384 (1996)

\bibitem{lus} J. G. Lussier et. al. J. Phys. Cond. Matter {\bf 7}, L325 
(1996)
\bibitem{ren} J. P. Renard et al. Europhys. Lett {\bf 30} 475 (1995) .

\bibitem{buch} J. P. Boucher and L. P. Regnault J. Phys. I {\it France} 
{\bf 6} 1, (1996).

\bibitem{cross} M. C. Cross and D. S. Fisher, Phys. Rev. B. {\bf 19} 402 (1979).

\bibitem{fuji} M. Fujita and K. Machida, J. Phys. Soc. J. Phys. Soc. Jpn. 
{\bf 53} 4395 (1984).   

\bibitem{nak} T. Nakano and H. Fukuyama J. Phys. Soc. Jpn. {\bf 49} 1679 
(1980); ibid {\bf 50} 2489 (1981).

\bibitem{got} G. S. Uhrig and H. J. Shultz, Phys. Rev. B {\bf 54} R9624 (1996);
G. S. Uhrig, Phys. Rev. Lett {\bf 79} 163 (1997).

\bibitem{cas} G. Castilla, S. Chakravarty, and V. J. Emery, Phys. Rev. Lett.
{\bf 75} 1823 (1995) 

\bibitem{rie} J. Riera and A. Dorby, Phys. Rev. B. {\bf 51} 16098 (1995).

\bibitem{bou1} G. Bouzerar, A. P. Kamf and F. Sch$\ddot{o}$nfeld, preprint
cond-mat/9701176.

\bibitem{bou2} G. Bouzerar, A. P. Kamf and G. I. Japaridze  accepted 
for publication in Phys. Rev. B. (cond-mat/9801046)

\bibitem{car} L. G. Caron and S. Moukouri, Phys. Rev. Lett {\bf 76} 4050 (1996)

\bibitem{uh} G. S. Uhrig, preprint, cond-mat/9801185

\bibitem{reg} L. P. Regnault et. al., Phys. Rev. B {\bf 5579} (1996)

\bibitem{bar} M. Braden, G. Wilkendorf, J. Lorenzana, M. Ain, G. J. Mcintyre
, M. Behruzi, G. Hegar, G. Dhalenne and A. Revcolvschi, Phys. Rev. B 
{\bf 54} 1105 (1996).   

\bibitem{frad} E. Fradkin and J. E. Hirsch Phys. Rev. B. {\bf 27} 1680 (1983).

\bibitem{su} W. P. Su, J. R. Schrieffer and A. J. Heeger Phys. Rev. B. {\bf 22}
2099 (1980).

\bibitem{cam}  D. K. Campbell and A. R. Bishop Nucl. Phys. B. 
{\bf 200} 297 (1982).

\bibitem{bray} J. W. Bray et. al. Phys. Rev. Lett. {\bf 11} 744 (1975).

\bibitem{jor} P. Jordan and E. Wigner, Z. Phys. {\bf 47} 631 (1928).

\bibitem{lang} I. Lang and Y. A. Firsov Sov. Phys. JETP {\bf 16} 1301 (1963).

\bibitem{das} A. N. Das and S. Sil, J. Phys. Cond. Matt. {\bf 5} 8265 (1993). 
\end{references}
\end{document}